\documentclass[a4paper, 12pt, fleqn]{article} 
\pdfoutput=1 
\usepackage{myarticle}

\mytitle{Intergroup Bias in Attitudes Toward Restrictions on Uncivil Political Expression and Its Underlying Mechanisms}
\mytitlenote{This working paper was uploaded on August 30, 2023. Note that it has not yet passed peer-review.}
\myname{Kohei Nishi}
\mynamenote{Research Fellow of Japan Society for the Promotion of Science and Ph.D. Student at the Division of Law and Political Science, the Graduate School of Law, Kobe University, Japan}

\myabstract{There appears to be a dilemma between the freedom of expression and protection from the adverse effects of uncivil political expression online. While previous studies have revealed various factors that affect attitudes toward freedom of expression and speech restrictions, it is less clear whether people have intergroup biases when forming these attitudes. To address this gap, the present study conducted a pre-registered online survey experiment and investigated people's attitudes toward uncivil political expression by randomizing its in-group and out-group affiliations.
The results revealed that people tend to perceive uncivil political expression directed from an out-group toward an in-group as more uncivil, compared to the expression originating from an in-group toward an out-group.
This difference subsequently influences their inclination to endorse speech restrictions when faced with uncivil political comments: stronger support for restrictions on expressions from the out-group toward the in-group as opposed to those from the in-group toward the out-group.
These findings should serve as a wake-up call to public opinion that advocates for restrictions on uncivil political expression.
}

\begindocument

\section*{Introduction}
Freedom of expression is a fundamental human right in democratic societies, as free access to discussion forums is essential for citizens to participate in politics.
However, expressions can sometimes hurt people's sentiments.
Today, social media and comment sections on news sites are filled with uncivil expressions against various communities \parencite{Coe2014, Kenski2017, Theocharis2020}.\footnote{The present study defines political incivility as ``a disrespectful or insulting expression that attacks an individual or group in political communication.''}
Thus, a dilemma exists between the importance of freedom of expression in a democracy and the dignity of those who are offended by uncivil expressions.

Previous studies have demonstrated that attitudes toward freedom of expression and speech restrictions depend on countries and individuals.
For example, in terms of inter-country differences, \textcite{Riedl2021} revealed that Germans than Americans are more prone to consider that law enforcement and an online platform bear the responsibilities of implementing measures against uncivil online comments.
Regarding gender differences, it was found that men tend to show more robust support than women for freedom of expression and less for speech restrictions \parencite{Downs2012, Lambe2004}.
Furthermore, regarding personal values, those with individualistic values are more supportive of freedom of expression, while those with right-wing authoritarianism beliefs are less so \parencite{Downs2012}.
However, while these studies have revealed the relationships between the characteristics of such countries or individuals and their attitudes toward freedom of expression or speech restrictions, there remains an ambiguity regarding the extent to which cognitive biases in intergroup relations exert an influence on these attitudes.

To address this gap in the literature, the present study hypothesized that people support speech restrictions more strongly when exposed to incivility by out-group members than by in-group members.
To evaluate this hypothesis, a pre-registered online survey experiment using a randomized controlled trial (RCT) approach was conducted on a sample of Japanese adults, and the results provided supporting evidence for it.

\section*{Bias in Intergroup Relations}
Over the past several decades, many social psychologists have investigated intergroup bias. It can be divided into two concepts, in-group favoritism and out-group derogation \parencite[see][]{Hewstone2002}.

In-group favoritism refers to individuals' tendency to show positive attitudes toward members of their own group (in-group). In-group favoritism emerges as a robust phenomenon, persisting even within the confines of the minimal group paradigm---a setting where individuals are assigned to practically inconsequential groups within laboratory settings \parencite{Doise1972, Hertel2001, Tajfel1971}. The underpinning framework for comprehending this tendency lies within the realm of social identity theory \parencite{Tajfel1979}.

Out-group derogation refers to their tendency to show negative attitudes toward members of the groups to which they do not belong (out-group). Out-group derogation is conceptually distinct from in-group favoritism \parencite{Hewstone2002}. The distinction between them has been shown in some empirical studies \parencite{Gibson2006, Mummendey1992}.

Intergroup bias is broadly used to explain social and political phenomena. For example, \textcite{Nicholson2012} conducted an experiment during the 2008 United States presidential election and found that people show more opposition to a policy when their out-group party leader supports the policy than when there are no such cues. Besides, \textcite{Simas2020} reported that individuals' level of empathy disposition is positively associated with favorable attitudes toward their in-group but negatively toward their out-group. The authors argued that such tendencies might contribute to political polarization. Intergroup bias also appears in people's perceptions of environmental issues. For example, \textcite{Jang2013} discovered that those who are informed about their own country's (i.e., their in-group's) over-consumption of energy attribute climate change to natural causes rather than human causes than those who are informed about another country's (i.e., the out-group's) over-consumption. Thus, intergroup bias is crucial for understanding social and political phenomena.

\section*{Intergroup Bias in Attitudes toward Restrictions of Political Expression}
The perspective of intergroup bias is also critical when studying people's attitudes toward speech restrictions. There is a possibility that governments abuse public opinion that is influenced by intergroup bias to suppress critical voices against themselves. Therefore, investigating people's attitudes toward restrictions of uncivil political expression from the perspective of intergroup bias is crucial to maintaining liberal democracy.

Although few such studies exist in the literature, \textcite{Lindner2009} provides related findings.
It examined whether the alignment between individuals' ideological group memberships and the expresser's affiliation influences their inclination to defend an extreme political expression.
The study employed an experiment involving U.S. citizens, wherein participants were randomly exposed to either a liberal or conservative extreme political expression.
The experimental results showed that liberal participants were more willing to protect the liberal extreme expression as opposed to the conservative one. On the other hand, conservative participants showed approximately the same level of willingness to protect the liberal and conservative expressions \parencite[see Figure 1 in][]{Lindner2009}.

While \textcite{Lindner2009} made excellent contributions to the literature,  its scope of interest slightly differs from that of the present study, and thus it does not directly answer the present study's question. More specifically, the outcome measurement in the previous study mixed and combined two concepts: (i) perceived level of harm of the expressions, and (ii) attitudes toward protecting/restricting the expressions. Conversely, the present study treats these two concepts separately to provide more nuanced knowledge. In addition, the previous study focused on offline rather than online political expression. However, the present study is instead interested in online uncivil political expression, which has become a more significant issue in recent years as the internet has become more widespread. The present study aimed to fill these gaps.

\section*{Hypotheses}
Based on the findings of previous studies in intergroup relations, this section introduces hypotheses about how people perceive uncivil political expression and the speech restrictions against them.

In terms of in-group favoritism, people are likely to perceive political expression by in-group members as more civil and worthy of protection as compared to expressions by out-group members. Thus, when people see political expression by in-group members, they are expected to proactively protect them in the public discourse forum.

In terms of out-group derogation and aggression, people are expected to perceive the expressions by out-group members as less civil and less worthy of protection and thus proactively support excluding them from the discourse space.
Particularly, when people are exposed to uncivil political expressions from the out-group directed at the in-group, they are likely to perceive harm originating from the out-group.
This perception is expected to arouse anger toward the out-group \parencite{Batson2009, Gordijn2001}, which in turn leads to aggressive attitudes against it \parencite{Anderson2002, Spanovic2010}.

Summarizing the above, the present study's argument is described in Figure \ref{mechanism}.
When individuals are exposed to uncivil political expressions from out-groups, their perception of the expression's incivility tends to be more pronounced compared to when they encounter similar expressions from in-groups.
This heightened perception subsequently contributes to a greater inclination to endorse restrictions on such expressions.

\vspace{10mm}
\begin{figure}[H]
\centering
\begin{tikzpicture}
    \tikzset{Start/.style={rectangle,  draw,  text centered, text width=4.7cm, minimum height=2.5cm}};
    \tikzset{Mid/.style={rectangle,  draw,  text centered, text width=4.7cm, minimum height=2.5cm}};
    \tikzset{End/.style={rectangle,  draw,  text centered, text width=4.7cm, minimum height=2.5cm}};
    \node[Start](a)at (0,0){Exposure to uncivil political expression by out-groups (compared to that by in-groups)};
    \node[Mid](b)at (6,0){Perceiving the expression as more uncivil};
    \node[End](c)at (12,0){Stronger support for restriction against the expression};
    \draw[-{Stealth[length=3mm]}]  (a) --(b);
    \draw[-{Stealth[length=3mm]}]  (b) --(c);
\end{tikzpicture}
\caption{Mechanism}\label{mechanism}
\end{figure}
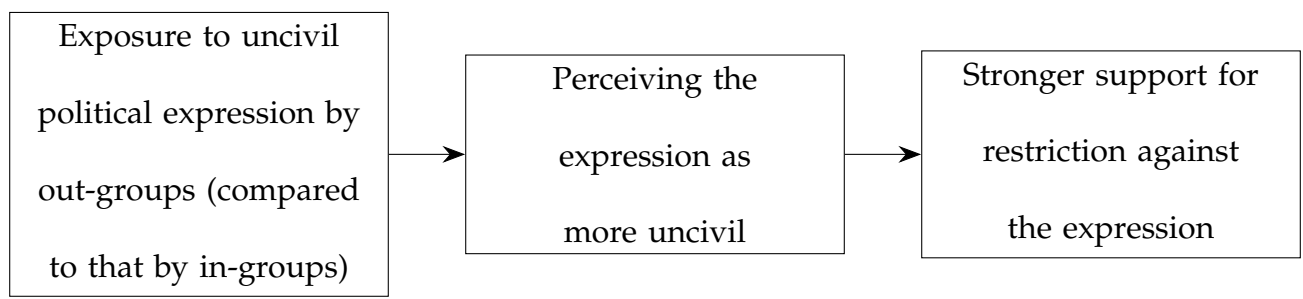

\textcite{Muddiman2017} found evidence that partially supports the above arguments, revealing that people perceive out-group political figures as more uncivil than in-group ones. However, it remains unclear whether a bias in perception of the incivility level can be extended to a bias in attitudes toward speech restrictions. Accordingly, the present study examines the following three hypotheses:

\begin{quote}
\textbf{Hypothesis 1}: People perceive uncivil political expression by out-group toward in-group as more uncivil than that by in-group toward out-group.
\end{quote}
\begin{quote}
\textbf{Hypothesis 2}: People support speech restrictions more strongly against uncivil political expression by out-group toward in-group than against that by in-group toward out-group.
\end{quote}
\begin{quote}
\textbf{Hypothesis 3}: The effect of exposure to uncivil political expression by out-groups (compared to that by in-groups) on the level of support for speech restrictions is mediated by the perceived level of incivility.
\end{quote}

Hypothesis 2 is the primary hypothesis of the present study, and Hypotheses 1 and 3 help to uncover the mechanism of Hypothesis 2.

\section*{Methods}
To test the above hypotheses, the present study conducted a pre-registered online survey experiment using an RCT approach. The preregistration document is available at Open Science Framework (\url{https://osf.io/zjtue}). After review and approval by the research ethics committee (IRB) of the author's institution,\footnote{Research Ethics Committee of Graduate School of Law, Kobe University (approval ID: 030010)} the survey was conducted between January 23 and February 1, 2023.\footnote{The present study conducted experiments twice after preregistration. Although the author believes that it is desirable to publish experimental data for open science practices, they inadvertently did not obtain consent from the participants about publishing data in the first experiment. Thus, experiment was conducted again with new participants, and consent was properly obtained. Therefore, the first experiment was positioned as a pilot study, and the second experiment was considered the main study. The latter's results are reported in this paper. The results of the two experiments show mostly similar tendencies regarding Hypotheses 1-3.}

\subsection*{Data Collection}
The participants of the survey were Japanese adults aged 18 to 70 recruited via Yahoo! Crowdsourcing, which is commonly used in other studies that are based in Japan \parencite[e.g.,][]{Ishii2023, Takizawa2022}.
The target number of participants was 2,500 as described in the preregistration document. A simplified quota-sampling approach was employed to collect a sample similar to the Japanese population, where 10 strata based on age and gender identity were set up and 250 participants were collected for each stratum.\footnote{Although the allocation of 250 participants per stratum does not precisely mirror Japan's population composition, it does maintain a general alignment with it. This approach was adopted because the target participant count can only be configured in increments of 50 in Yahoo! Crowdsourcing.} Following recruitment, participants were directed to the web-based survey platform, Qualtrics, wherein the survey was administered. Responses surpassing the intended count were logged, and a subset of these responses was subsequently excluded. This encompassed participants who disagreed with the data utilization after the debriefing, those with duplicated IP addresses, individuals who did not pass an attention check question, and those who provided inappropriate age responses.\footnote{An attention check was conducted with the Directed Questions Scale approach \parencite{Maniaci2014} to detect satisficers. More specifically, before the treatment, the following four items were presented: ``I support an amendment of the Japanese Constitution's Article 9,'' ``I support the legal recognition of selective dual surname system for married couples,'' ``Please select `I do not think so' for this item,'' and ``I support the legal recognition of same-sex marriage.'' The set of answer labels for these items was 1 = \textit{I do not think so at all}, 2 = \textit{I do not think so}, 3 = \textit{Neither/nor}, 4 = \textit{I think so}, 5 = \textit{I strongly think so}, and 6 = \textit{I do not know or prefer not to answer}. The participants who did not select \textit{I do not think so} for the third item were considered not to have carefully read the question text and thus considered satisficers.} Following these exclusions, the sample size amounted to \textit{N} = 2,581.\footnote{The number of participants exceeded the target for technical reasons.}



\subsection*{Measurement of Pretreatment Variables}
Before the treatments, the participants were asked about their demographic information (gender, age, education level, and income level). These variables were used to ensure that the randomization was successful. The results of the balance check using these variables are presented in the Appendix section.

In addition, the participants were asked about their feelings toward the Kishida Cabinet, the incumbent Japanese cabinet at the time of the survey, using a feeling thermometer scale. The scale ranged from 0 to 100, where 0 means \textit{strong antipathy}, 50 means \textit{neutral}, and 100 means \textit{strong favorability}.\footnote{The survey experiment was conducted in Japanese. The survey questions presented in this paper are translated versions.} Thus, participants who scored less than 50 were classified as anti-Kishida, while those who scored greater than 50 were classified as pro-Kishida.
Participants who responded with a score of 50 to this question, as well as those who declined to answer, could not be definitively categorized as either pro-Kishida or anti-Kishida. As a result, they were omitted from the sample (\textit{N} = 609 out of 2,581).
Hence, the sample size for the main statistical analysis was \textit{N} = 1,972.

\subsection*{Treatment}
The participants were then presented with one of two randomly selected treatments: Treatment A or B shown in Figure \ref{treatments}. The randomization was implemented using Qualtrics's randomizer. The treatments are designed to resemble comments in the comment section of a news site. Treatment A is an uncivil comment from an anti-Kishida individual toward a pro-Kishida community: ``Japan is being destroyed by the idiots who support the Kishida Cabinet.'' In contrast, Treatment B is an uncivil comment from a pro-Kishida individual toward an anti-Kishida community: ``Japan is being destroyed by the idiots who criticize the Kishida Cabinet.''

Based on the treatments, participants were divided into ``incivility by out-group'' and ``incivility by in-group'' conditions (\textit{N} = 1,029 and \textit{N} = 943, respectively). The anti-Kishida participants assigned to Treatment A were classified as ``incivility by in-group.'' In contrast, those assigned to Treatment B were classified as ``incivility by
out-group.'' Furthermore, the pro-Kishida participants assigned to Treatment A were classified as ``incivility by out-group.'' In contrast, those assigned to Treatment B were classified as ``incivility by in-group.''

\vspace{5mm}
\begin{figure}[H]
  \centering
  \begin{minipage}[t]{0.42\linewidth}
    \centering
    \includegraphics[keepaspectratio, scale=0.6]{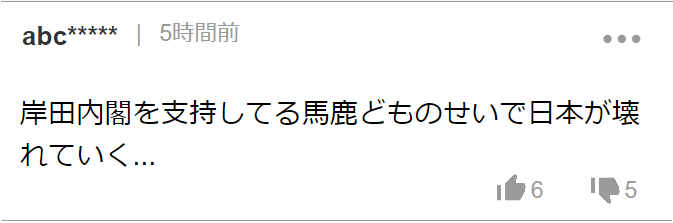}
    \caption*{\tablenotesize Treatment A: uncivil comment from an anti-Kishida individual toward a pro-Kishida community}
  \end{minipage}
  \hspace{0.05\linewidth}
  \begin{minipage}[t]{0.42\linewidth}
    \centering
    \includegraphics[keepaspectratio, scale=0.6]{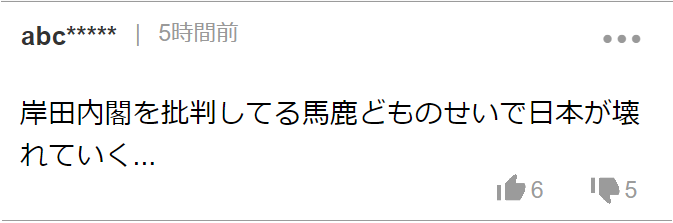}
    \caption*{\tablenotesize Treatment B: uncivil comment from a pro-Kishida individual toward an anti-Kishida community}
  \end{minipage}
  \caption{Treatments}\label{treatments}
\end{figure}

\subsection*{Measurement of Outcome Variables}
To measure the participants' perceived level of incivility of the presented comment, they were asked, ``Is this comment an uncivil expression against others?'' using a 7-point Likert-type scale (1 = \textit{I do not think so at all}, 2 = \textit{I do not think so}, 3 = \textit{I rather do not think so}, 4 = \textit{Neither/nor}, 5 = \textit{I rather think so}, 6 = \textit{I think so}, 7 = \textit{I strongly think so}; the option ``I do not know or prefer not to answer'' was also available). Subsequently, to measure their level of support for a restriction against the presented comment, they were asked, ``Should this comment be deleted by the website administrator?'' using the same 7-point scale.

\subsection*{Statistical Analysis}
To test Hypotheses 1 and 2, mean values of the outcome variables (on a 7-point scale) for the two conditions were calculated. Following this, two-tailed Welch's \textit{t}-tests were conducted on the mean differences in outcomes between the two conditions at the .05 level of significance. They were performed using a pooled sample consisting of both anti-Kishida and pro-Kishida participants as pre-registered.

To test Hypothesis 3, a causal mediation analysis \parencite{Imai2011} was conducted to estimate the average causal mediation effect (ACME).\footnote{The mediation package in R \parencite{mediation} was used to conduct causal mediation analysis. Both the mediator model and outcome model were estimated with the ordered probit method.} In the analysis, the treatment is exposure to uncivil political expression by out-groups (with that by in-groups as a reference category), the outcome is the level of support for speech restrictions, and the mediator is the perceived level of incivility. Age, gender, education level, and income level were controlled. Again the analysis was performed using the pooled sample.

Some participants answered ``I do not know or prefer not to answer'' for the outcome questions (\textit{N} = 15 for the perceived level of incivility question and \textit{N} = 16 for the support for restrictions question). Such missing values were processed by multiple imputation with the EMB algorithm \parencite{Honaker2011}.\footnote{\textit{M} = 1,000 imputed data sets were generated by Amelia II package in R \parencite{Honaker2011}. Then, in the analysis for Hypotheses 1 and 2, the imputated data sets were analyzed separately and the results were combined using mi.t.test() function in the MKmisc package in R \parencite{mkmisc}. In the analysis for Hypothesis 3, ACMEs were calculated for each imputated data set using mediation package in R, and the results were combined by averaging them. Confidence intervals of the ACMEs were calculated by percentile method. A model for predicting missing values encompassed several variables, including the experimental condition (as represented by a dummy indicator for the "incivility by out-group" condition), age, gender identity, education level, and income level.}

\subsection*{Advantage as a Test Case}
The Japanese sample has an advantage as a test case. According to a previous study, political polarization in Japan is much more moderate than in the United States \parencite{Rojas2019}. Thus the Japanese people are less likely to show political intergroup bias as compared with other countries that are highly polarized, like the United States. Thus, the Japanese sample serves as a scenario wherein the Hypotheses were less likely to find support, consequently furnishing robust evidence.

\section*{Results}
\vspace{-10mm}
\subsection*{Balance Check}
Table \ref{balance} in the Appendix section presents the balance of pretreatment variables, where the percentages of participants who fall into each category are presented for each condition. As the table shows, no substantial differences were found between the pretreatment variables in the two conditions. Thus, the randomization seems to have been successful.

\subsection*{Results for Hypotheses 1 and 2}
Figure \ref{results_h1&2} displays the main results. The sizes of the bars represent the means for each condition, and the lines on the bars are their 95\% confidence intervals.

The left panel in Figure \ref{results_h1&2} shows the results for Hypothesis 1. Consistent with the expectation, the participants perceived an uncivil comment by an out-group member as more uncivil than that by an in-group member. More specifically, the mean perceived level of incivility of the participants exposed to an uncivil comment from the out-group was 5.07 with 95\% CI [4.97, 5.18] on a 7-point scale, while for those exposed to an uncivil comment from the in-group, the mean was 4.83 with 95\% CI [4.73, 4.94]. The mean difference between the conditions was 0.24, which was statistically significant (\textit{t}(1,953.40) = 3.12, \textit{p} = .002). This finding supports Hypothesis 1.

The right panel in Figure \ref{results_h1&2} shows the results for Hypothesis 2. Again, consistent with the expectation, the participants supported the speech restriction more strongly when exposed to an uncivil comment by an out-group member than by an in-group member. More specifically, the mean support for speech restriction of the participants exposed to an uncivil comment from the out-group was 4.06 with 95\% CI [3.95, 4.17]. In contrast, for those exposed to an uncivil comment from the in-group, the mean was 3.60 with 95\% CI [3.49, 3.70]. The mean difference between these two conditions was 0.47, which was statistically significant (\textit{t}(1,950.93) = 5.97, \textit{p} < .001). This finding supports Hypothesis 2.

\vspace{5mm}
\begin{figure}[H]
    \centering
    \hspace{-10mm}
    \includegraphics[scale=0.35]{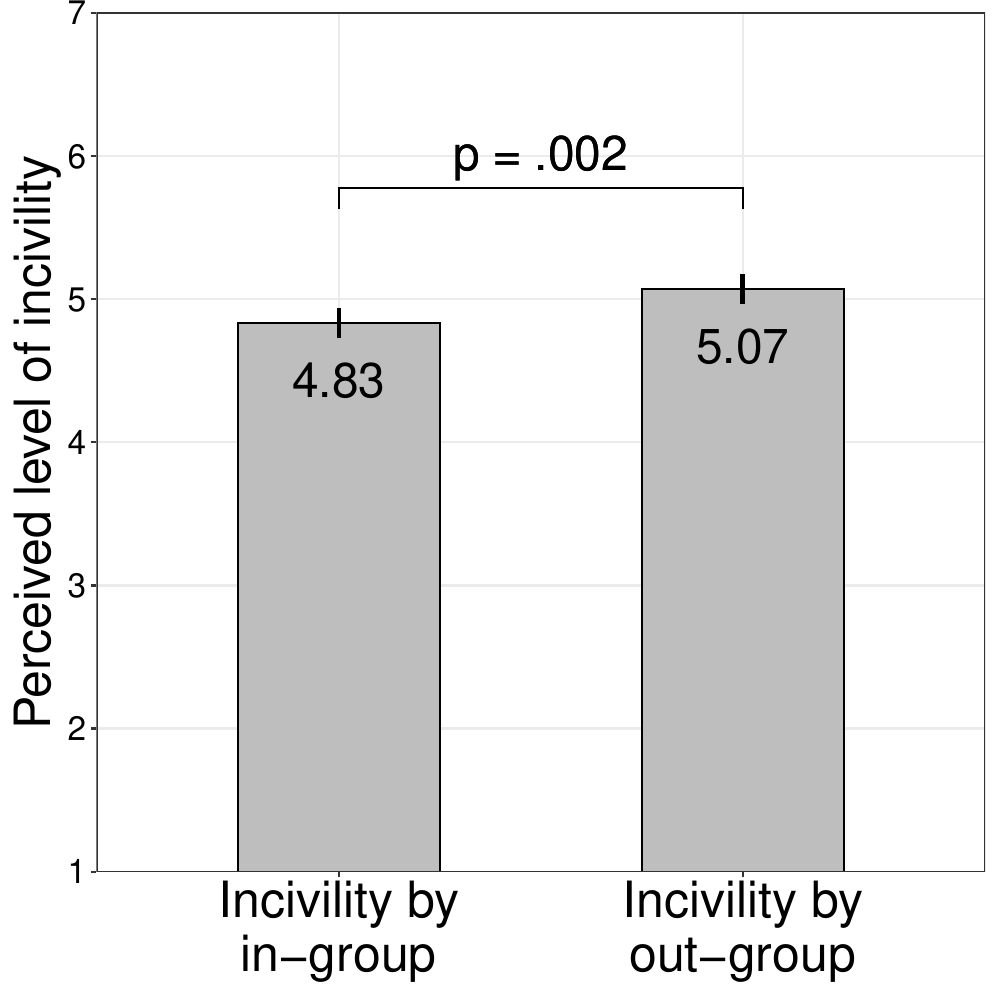}
    \hspace{5mm}
    \includegraphics[scale=0.35]{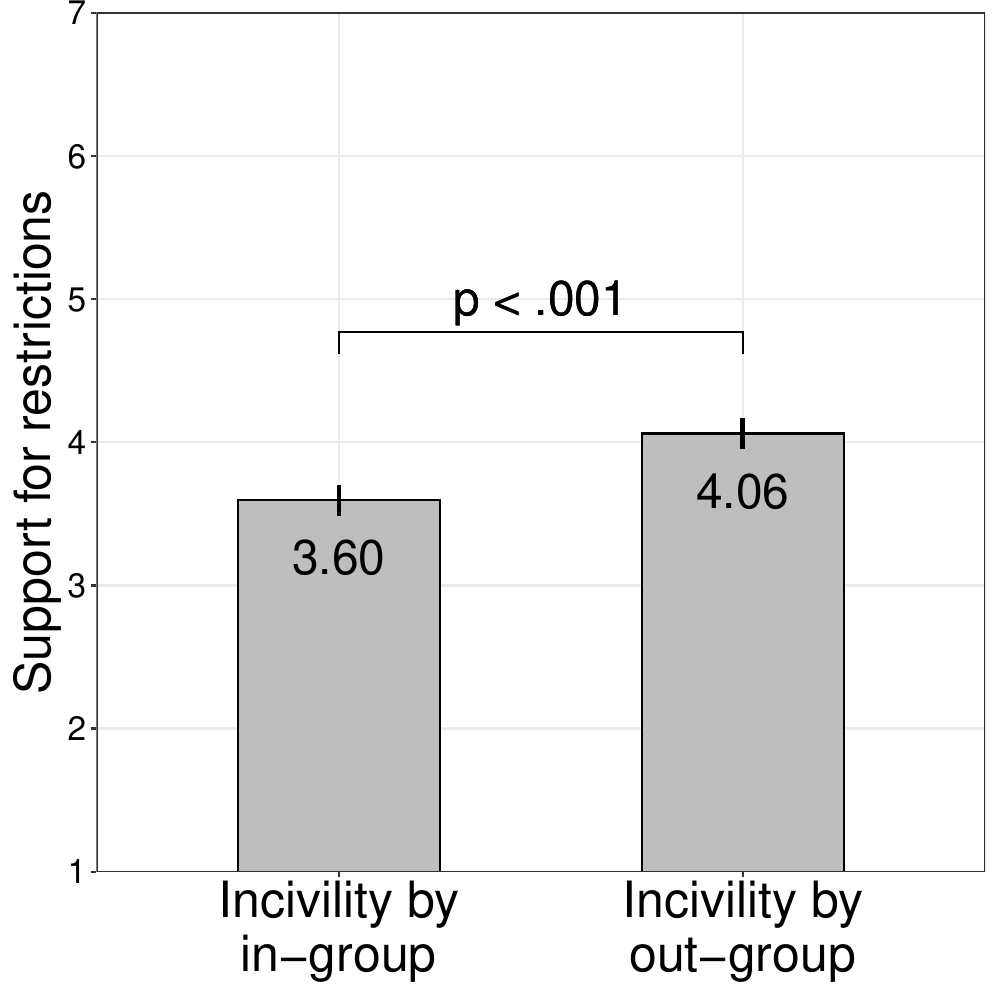}
    \hspace{-10mm}
    \caption{Results for Hypotheses 1 and 2}
    \label{results_h1&2}
\end{figure}

\subsection*{Results for Hypothesis 3}
Figure \ref{results_h3} illustrates the results of the causal mediation analysis conducted for Hypothesis 3. This hypothesis anticipates that exposure to uncivil political expression from out-groups (compared to that from in-groups) intensifies the endorsement of restrictions against the expressed content. This heightened support is believed to be facilitated by the perceived level of incivility associated with the expression.

The horizontal axis indicates the categories of the outcome variable (the level of support for restrictions), and the vertical axis indicates the magnitude of the ACME on the probability that each category is selected. The plots represent the estimates of the ACMEs and the lines represent their 95 percentile confidence intervals. As shown, the estimates of the ACMEs are negative for categories one to three (1 = \textit{I do not think so at all} to 3 = \textit{I rather do not think so}), indistinguishable from 0 for category four (4 = \textit{Neither/nor}), and positive for categories five to seven (5 = \textit{I rather think so} to 7 = \textit{I strongly think so}). This implies the presence of a mechanism wherein the distinction in the originator of the uncivil comment (in-group or out-group) leads to the difference in the perceived level of incivility associated with the uncivil comment, which in turn leads to the difference in the likelihood of supporting the restriction of the uncivil comment. This result supports hypothesis 3.

\vspace{5mm}
\begin{figure}[H]
    \centering
    \includegraphics[scale=0.34]{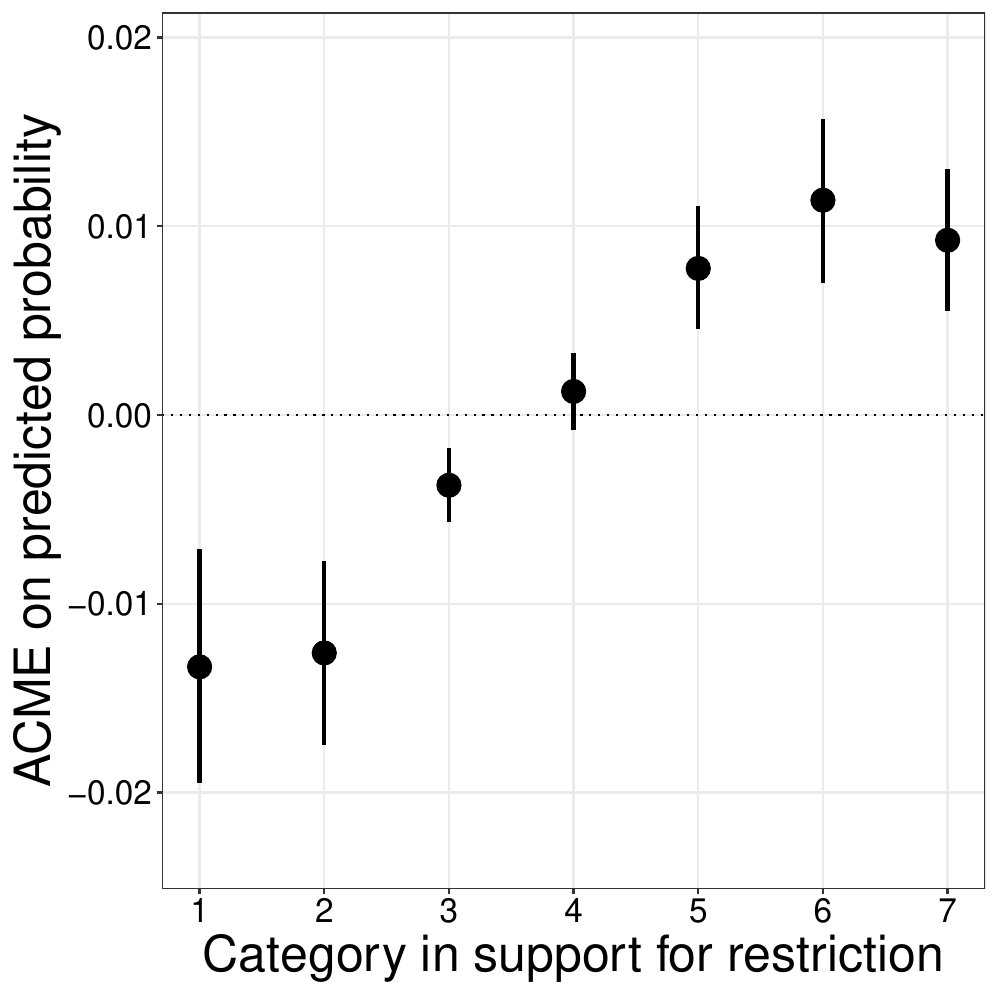}
    \caption{Results for Hypothesis 3}
    \label{results_h3}
    \vspace{-10mm}
\end{figure}

\subsection*{Results for Additional Analysis}
Figures \ref{anti} and \ref{pro} show the results for additional analysis, which was conducted on the anti-Kishida sample (\textit{N} = 1,613) and the pro-Kishida sample (\textit{N} = 359) separately.

As shown in Figure \ref{anti}, results for the anti-Kishida sample show similar tendencies as that of the main analysis. Both the mean perceived level of incivility and the mean support for restrictions were greater in ``incivility by out-group'' condition than in ``incivility by in-group'' condition, with the mean difference 0.23 and 0.48 respectively, which were both statistically significant at the \textit{p} < .05 level (\textit{t}(1,596.64) = 2.64, \textit{p} = .008, and \textit{t}(1,591.48) = 5.57, \textit{p} < .001, respectively).

\vspace{5mm}
\begin{figure}[H]
    \centering
    \hspace{-10mm}
    \includegraphics[scale=0.35]{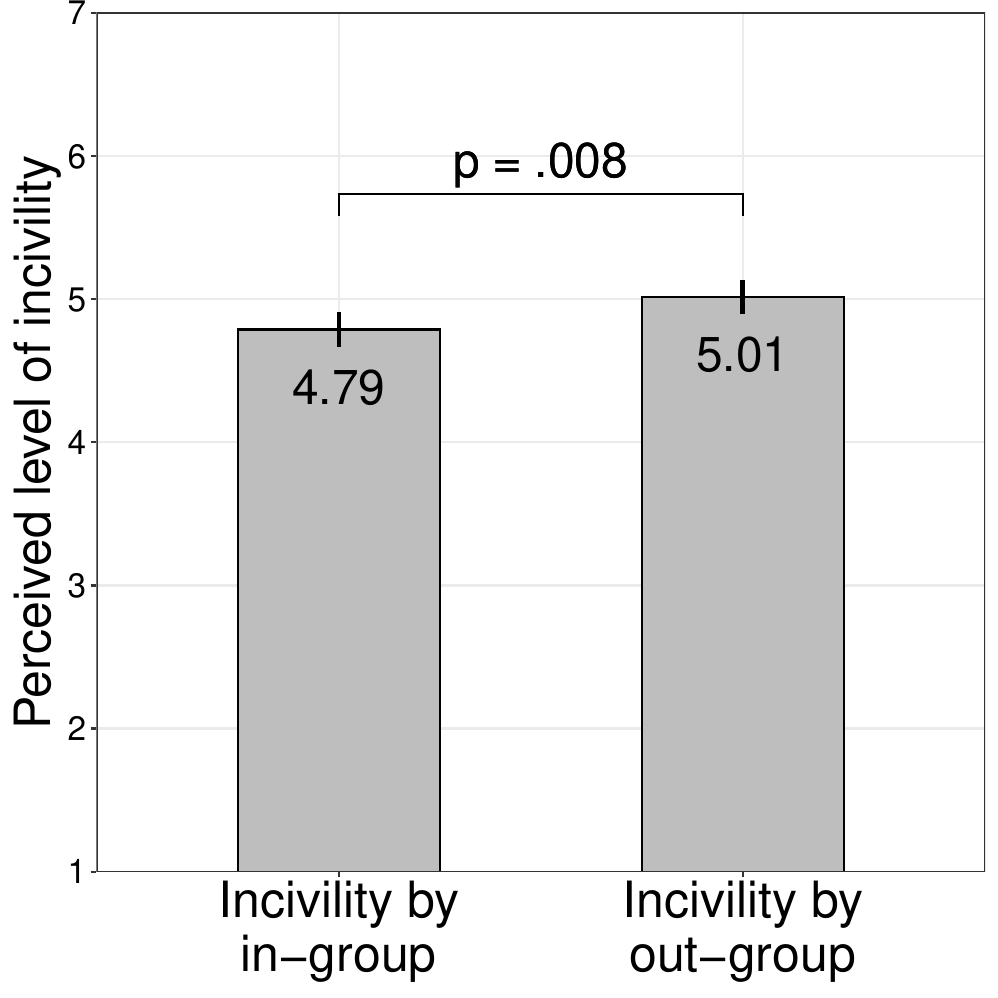}
    \hspace{5mm}
    \includegraphics[scale=0.35]{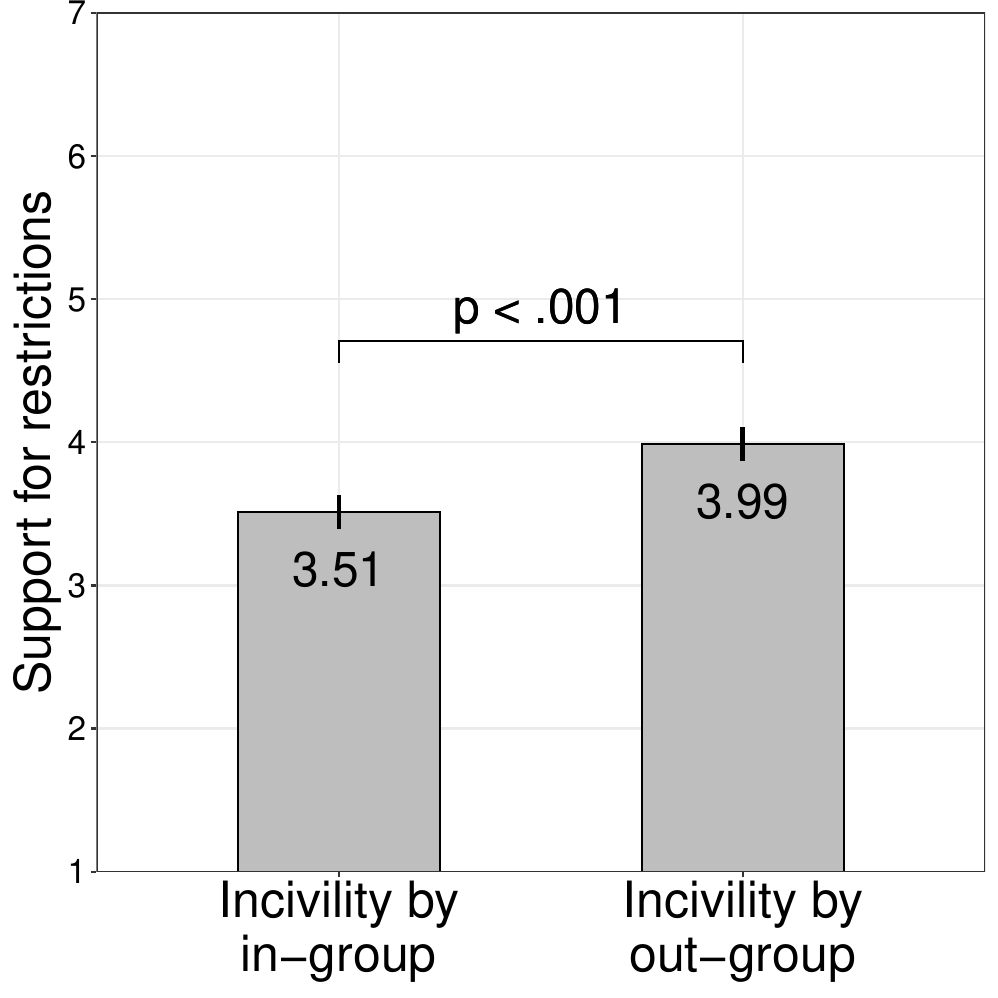}
    \hspace{-5mm}
    \caption{Results of the Additional Analysis on the Anti-Kishida Sub-Sample}
    \label{anti}
\end{figure}

Results for the pro-Kishida sample are more ambiguous. As shown in Figure \ref{pro}, point estimates of both outcomes were greater in the ``incivility by out-group'' condition than in the ``incivility by in-group'' condition, with the mean difference being 0.29 and 0.40 respectively. The former was statistically not significant at the \textit{p} < .05 level (\textit{t}(352.47) = 1.73, \textit{p} = .08), while the sign of the difference was consistent with Hypothesis 1. The latter was statistically significant at the \textit{p} < .05 level (\textit{t}(353.83) = 2.21, \textit{p} = .03).

\vspace{5mm}
\begin{figure}[H]
    \centering
    \hspace{-10mm}
    \includegraphics[scale=0.35]{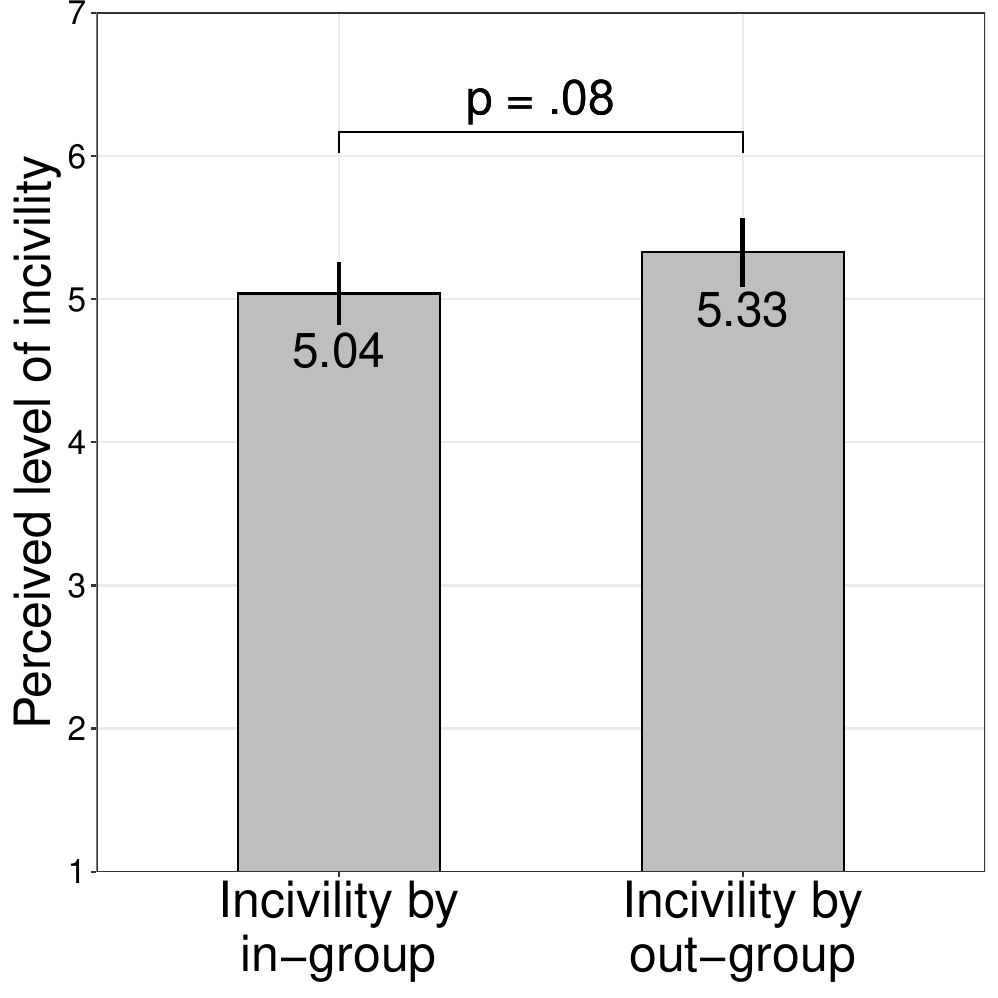}
    \hspace{5mm}
    \includegraphics[scale=0.35]{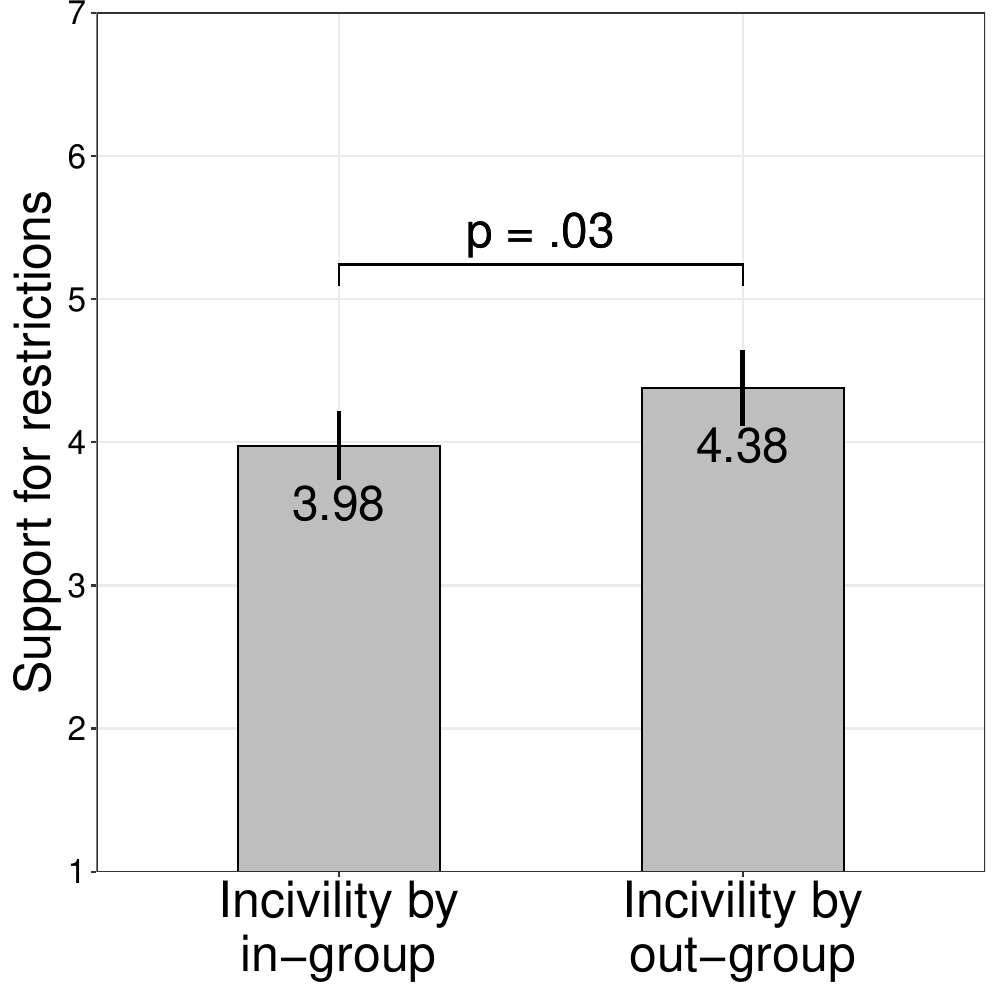}
    \hspace{-5mm}
    \caption{Results of Additional Analysis on Pro-Kishida Sub-Sample}
    \label{pro}
\end{figure}

In general, statistical non-significance does not necessarily imply the absence of differences. Also, these findings are just the results of an additional analysis. Therefore, they do not immediately shake the conclusion of the main analysis. However, it must be carefully determined whether the hypothesized tendencies are always found in both conflicting groups.

\section*{Discussion}
The present study investigated whether people have intergroup cognitive biases when forming attitudes toward uncivil political expression and restrictions against them. An online survey experiment on a Japanese sample found that people perceived uncivil political expression from the out-group toward the in-group as more uncivil than that from the in-group toward the out-group. Furthermore, it was found that this difference led people to be more supportive of speech restrictions against uncivil political comments from the out-group toward the in-group than the other way around.

The present study's findings should serve as a wake-up call to public opinion supporting restrictions on uncivil political expression. Although Japan is a liberal democratic country, many of its citizens appear to support restrictions on uncivil political expression. According to the present study's survey results, as many as 38.69\% of the participants responded in support of a restriction of the presented comments (but note that it is not from a complete nationally representative sample).\footnote{When calculating this percentage, participants who responded with scores ranging from 5 to 7 on the 7-point scale for the question about support for restrictions were categorized as advocates of speech restrictions. The sample encompassed not only participants with anti- and pro-Kishida stances but also those who maintained a neutral position, resulting in a total sample size of \textit{N} = 2,581. Missing values (\textit{N} = 28) were processed by the EMB multiple imputation with \textit{M} = 1,000 imputations.}
However, the survey results also suggest that people perceive uncivil political expression from out-groups as harsher and thus support restrictions against them.
This implies that it is plausible that individuals who generally advocate for restrictions on uncivil political expression may do so primarily by directing their attention toward uncivil expressions originating from the out-groups.
Therefore, if restrictions on uncivil political expression were to be strengthened by governments or website administrators based on such public opinions, it is likely that even the expressions by in-groups, which people thought should not be regulated, would be restricted.
Governments can abuse such intergroup biases to strengthen speech restrictions for their benefit, posing a potential crucial threat to liberal democracy.

Of course, the present study has several limitations.
First, there might be an issue of external validity. The survey in this study was conducted on a Japanese sample. Hence, it is unknown whether similar phenomena can be observed in regions other than Japan. However, as explained in the Methods section, the present study considers this sample a hard case. Thus, this issue is not severe.
Second, there also might be an issue with convenience sampling. The survey was conducted with the participants recruited via Yahoo! Crowdsourcing, a type of convenience sample. Hence, the sample might not perfectly represent the demographic composition of Japan's population. In addition, there may possibly be a selection bias; those highly interested in social and political topics might have been over-represented.
Third, within the experiment, participants were solicited to express their attitudes regarding the imposition of restrictions on particular uncivil expressions that were presented immediately prior to the inquiry, rather than conveying their stance on the restriction of uncivil political expression in a broader context.
Fourth, it is unclear whether the bias observed in the experiment is actually due to intergroup bias. Another possibility is that the bias might be caused by differences in the level of agreement or sympathy with the presented comments.
Therefore, future research is needed to overcome these limitations.

\theendnotes

\section*{Funding}
This work was partly supported by the JSPS KAKENHI Grant Number 22J21515 and the Research Fund from the Quantitative Methods for International Studies Program at Kobe University.

\myreference

\startappendix

\begin{table}[H]
\centering\tablesize
\caption{Balance of Pre-treatment Variables} 
\label{balance}
\begin{tabularx}{115mm}{lRR}
  &  \\ 
  \hline
 & \begin{tabular}{c}Incivility by\\in-group\end{tabular} & \begin{tabular}{c}Incivility by\\out-group\end{tabular} \\ 
   \hline
Gender: Man & 50.59\% & 52.83\% \\ 
  Gender: Woman & 49.30\% & 47.07\% \\ 
  Gender: Others &  0.11\% &  0.10\% \\ 
  Age: 18 to 29 years old & 16.86\% & 18.66\% \\ 
  Age: 30 to 39 years old & 18.03\% & 20.51\% \\ 
  Age: 40 to 49 years old & 22.16\% & 18.66\% \\ 
  Age: 50 to 59 years old & 20.47\% & 19.14\% \\ 
  Age: 60 to 70 years old & 22.48\% & 23.03\% \\ 
  Education: Junior high school &  1.91\% &  0.97\% \\ 
  Education: High school & 25.24\% & 29.15\% \\ 
  Education: Junior college, etc. & 22.80\% & 20.89\% \\ 
  Education: College & 45.07\% & 44.80\% \\ 
  Education: Graduate school &  4.98\% &  4.18\% \\ 
  Income: Less than 2 million yen & 19.50\% & 21.35\% \\ 
  Income: 2 to less than 4 million yen & 25.03\% & 25.14\% \\ 
  Income: 4 to less than 6 million yen & 21.26\% & 20.90\% \\ 
  Income: 6 to less than 8 million yen & 15.97\% & 12.86\% \\ 
  Income: 8 to less than 10 million yen &  9.31\% &  8.84\% \\ 
  Income: 10 million yen or more &  8.93\% & 10.91\% \\ 
   \hline
\end{tabularx}
\end{table}

\end{document}